\documentclass[aps,pra,showpacs,onecolumn]{revtex4}
\usepackage[english]{babel}
\usepackage[T1]{fontenc}
\usepackage{amsmath,amsfonts,amssymb,float,graphicx,epsfig,color,times,bbm}

\begin{document}
\author{J. P. Santos}
\affiliation{Centro de Ci\^encias Naturais e Humanas, Universidade Federal do ABC, Santo Andr\'e, 09210-170 S\~ao Paulo, Brazil}
\title{Entanglement in atomic systems interacting with single-photon pulses in free space}
\author{F. L. Semi\~ao}
\affiliation{Centro de Ci\^encias Naturais e Humanas, Universidade Federal do ABC, Santo Andr\'e, 09210-170 S\~ao Paulo, Brazil}
\begin{abstract}
New technologies providing tight focusing lens and mirrors with large numerical apertures and electro-optic modulation of single photons are now available for the investigation of photon-atom interactions without a cavity. From the theoretical side, new models must be developed which take into account the intrinsic open system aspect of the problem as well as details about atomic relative positions and temporal (spectral) features of the pulses. In this paper, we investigate the generation of nonclassical correlations between two atoms in free space interacting with a quantized pulse containing just one single photon. For this purpose, we develop a general theoretical approach which allows us to find the equations of motion for single- and two-body atomic operators from which we obtained the system density operator. We show that one single photon is capable of promoting substantial entanglement between two initially ground state atoms, and we analyze the dependence of these correlations on atomic relative positions (vacuum effects) and spectral features of the single photon pulse. 
\end{abstract}
\maketitle
\section{Introduction} 
The strong coupling between photons and atoms has been demonstrated in many different experiments in the context of cavity quantum electrodynamics (CQED). In one typical physical setting, Rydberg atoms with large electric dipoles couple very efficiently to a single photon enclosed in a high quality cavity sustaining a resonant electromagnetic field mode \cite{Haroche_Review}. These favorable conditions cause the atom to couple strongly to a particular mode of the quantized electromagnetic field in the cavity. In principle, this strong coupling between single photons and atoms could be used to implement distributed quantum communication (quantum networks) where localized information carriers (atoms) exchange information by means of flying qubits (photons) \cite{Quantum Networks}. However, the cavities used to coherently couple photons and atoms possess very long photon storage times \cite{Haroche_Review} making photon extraction inefficient. Even in an intermediate coupling regime where photons may easily leak the cavity and information could in principle be transferred to another node (cavity with atoms) \cite{Quantum Networks}, the scaling of the multi-cavity system is experimentally hard to achieve.

This problem has been motivating the investigation of possible experimental settings to promote the strong coupling of a single photon with atoms in free space, avoiding the complications related to the use of cavities for quantum networking \cite{Rist}. On one hand, there is the question of what kind of field mode in free space would offer the highest electric-field intensity at the atom for an incoming beam. This question has already been answered \cite{edw} and it turns out to be an eletric-dipole wave, i.e., the time reversed version of the wave originally emitted by an oscillating electric dipole at the point of interest (an atom). From this fact, it follows that in order to achieve the strongest atom-electromagnetic field coupling in free space, one should build an optical system (mirrors, lens, etc.) capable of creating incoming light waves with great overlap with those dipole waves. The application of deep parabolic mirrors and other optical settings for such a task has been reported \cite{parabolic}. 

From a quantum mechanical point of view, where the concept of photon is meaningful, one should ask about the existence of an initial state of the free space single-photon pulse capable of perfectly driving an atom from its ground state to an excited state (a $\pi-$pulse). In fact, this state exists and its properties are intrinsically related to the dipole wave mentioned above in the scope of classical electromagnetism \cite{perf_exc}. The single photon state capable of perfectly exciting the atom in free space is the time reversed photon state originating from spontaneous decay of a two-level system, and its temporal shape is that of a rising exponential \cite{perf_exc}. From the experimental side, electro-optic modulation of single photons provide the key ingredient to study the transient response of atoms to different single-photon waveforms \cite{el_op}. All these recent developments in addition to the experimental demonstration of strong coupling between weak coherent light and a single trapped atom without a cavity \cite{strong} make the investigation of models and effects related to single photon excitation of atoms in free space a very timely field of investigation. Following these lines, we present here a model to treat the interaction of two atoms with a quantized pulse in free space, and we study the power of a single photon to create entanglement between the atoms. 
It is important to remark that when two quantum subsystems interact with a commom bath, such as the vacuum electromagnetic field, they may become entangled depending strongly on their initial state \cite{DanielBraun}. In this sense, a classical laser field may be used to prepare a suitable initial state and then produce entanglement between the atoms. This is so because atom-atom interactions appear as a consequence of the free space vacuum environment, as seen previously in the context spontaneous emission of a collection of atoms in free space \cite{Natoms,Livro Ficek,Thiru}. This is a \emph{passive approach} to induce entanglement in the system. One may also  want to attempt an \emph{active approach} where the dynamics of entanglement may be somehow externally controlled. For a two-atom system in the presence of vacuum as a commom reservoir, a classical laser field can be used to controllably populate collective states (symmetric and antisymmetric) and build up a considerable amount of entanglement between the atoms \cite{Tanas}. In the present work, we extend this active control paradigm to an all quantized approach which, for the single photon level, must be more accurate than the approach considering a semiclassical model with an attenuated laser beam. Besides, by building a fully quantized model, one can explore quantum resources not present in classical electromagnetic pulses, such as light squeezing or entanglement to interact with the atomic system in free space.

\section{Model} 
The Hamiltonian for the system consisting of two two-level atoms sitting at positions ${\bf r}_1$ and ${\bf r}_2$ and interacting with the quantized pulse can be written (in the interaction picture and RWA approximation) as
\begin{eqnarray}\label{H}
H^I=-i\hbar\sum_\lambda\int d{\bf k}
[a^{}_{{\bf k},\lambda}(
g_{1;{{\bf k},\lambda}}\sigma_1^++g_{2;{{\bf k},\lambda}}\sigma_2^+) e^{-i(\omega_{k}-\omega)t}-{\rm{H.c.}}],
\end{eqnarray}
where $\sigma_{i}^+=|e\rangle_i\langle g|$ with $i=1,2$ are raising atomic operators. We consider here the case where the atomic transition frequencies of both atoms are equal and given by $\omega$. In the absence of any mode-selecting cavity, all modes of the electromagnetic radiation field couple to the atom, and the
coupling constant $g_{i;{{\bf k},\lambda}}$ is given by
\begin{align}\label{g}
g_{i;{{\bf k},\lambda}}=
\mbox{\boldmath$\mu$}_{i}\cdot{\mbox{\boldmath$\epsilon$}}_{{\bf k},\lambda}u_{{\bf k},\lambda}({\bf r}_{i})\sqrt{\frac{\omega_k}{(2\pi)^32\hbar\varepsilon_0}},
\end{align}
where $\mbox{\boldmath$\epsilon$}_{{\bf k},\lambda}$ are unit polarization vectors, $u_{{\bf k},\lambda}({\bf r})$ are the spatial mode functions and
$\mbox{\boldmath$\mu$}_{i}$ are electric dipole moments of the atoms.
%

In order to evaluate the quantum correlations driven by a single photon we need to obtain the dynamical evolution of the atomic density operator $\rho(t)$. We achieved that by we employing Heisenberg equations for treating this problem, generalizing the single atom approach discussed in \cite{Scarani}. In some respect, this approach can be seem as a form of input-output theory \cite{gc} which is largely known for providing tools for the treatment of general baths. The atomic density operator can be obtained from solutions of reduced Heisenberg equations for one- and two-body atomic operators having the field variables traced out. In the Schr\"{o}dinger picture, the density operator for two two-level atoms can be written as \cite{PLA_222_21}
\begin{eqnarray}\label{rho}
\rho(t)=\frac{1}{4}[
I\otimes I
+{\bf d}(t)\cdot\mbox{\boldmath$\sigma_1$}\otimes I
+I\otimes{\bf s}(t)\cdot\mbox{\boldmath$\sigma_2$}
+\sum_{n,m=1}^{3}O_{n,m}(t)\sigma_1^n\otimes\sigma_2^m],
\end{eqnarray}
where $I$ stands for the identity operator for each two level atom, $\mbox{\boldmath$\sigma_i$}=\{\sigma_i^x,\sigma_i^y,\sigma_i^z\}=\{\sigma_i^1,\sigma_i^2,\sigma_i^3\}$, and the coefficients are calculated as \textit{expectation values} $d_i={\rm Tr}(\rho\sigma_1^i)$, $s_i={\rm Tr}(\rho\sigma_2^i)$ and $O_{nm}={\rm Tr}(\rho\sigma_1^n\sigma_2^m)$. We may evaluate the expectation values using the Heisenberg picture and then plug the results in (\ref{rho}) to obtain the density operator in Schr\"{o}dinger picture.

We then aim now to solve the relevant set of Heisenberg equations for the whole system (atoms$+$pulse) prepared in the initial state $|\Psi\rangle=|g_1,g_2,1_p\rangle$, where the single photon state reads \cite{Scully}
\begin{align}\label{1p}
|1_p\rangle=\sum_\lambda\int d{\bf k} g^*_{0;{{\bf k},\lambda}}f(\omega_k)a_{{\bf k},\lambda}^\dag|0\rangle,
\end{align}
and the atoms are initially in the ground state. The factor $g_{0;{{\bf k},\lambda}}$ is given by (\ref{g}) and evaluated at the position ${\bf r_0}$ representing the focal point. This corresponds, for example, to the field state after spontaneous emission by an atom at ${\bf r_0}$ or an incoming wave focused on an fictional atom at ${\bf r_0}$. It all depends on the choice of $f(\omega_k)$ which is linked to the temporal features of the pulse by 
\begin{align}\label{t}
\xi(t)=\frac{\sqrt{\gamma_0}}{2\pi}\int d\omega_k f(\omega_k)e^{-i(\omega_k-\omega)t}.
\end{align}
We will treat here two representative single-photon waveforms that can be readily generated using electro-optic modulation \cite{el_op}, namely the rising exponential pulse and the Gaussian pulse. They respectively correspond to the choices
\begin{eqnarray}\label{r}
\xi(t)= \left\{
\begin{array}{lcl}
\sqrt{\Omega}\exp(\Omega t/2) & {\rm for} & t<0\\
0 & {\rm for} & t>0\\
\end{array}
\right.,
\end{eqnarray}
and
\begin{eqnarray}\label{ga}
\xi(t)=\left(\frac{\Omega^2}{2\pi}\right)^{1/4}\exp(-\Omega^2t^2/4).
\end{eqnarray}

From now on we denote $g_{i;{{\bf k},\lambda}}$ as $g_i$ but the dependence on polarization and wave vector must be properly taken into account in calculations using these quantities. With that initial state, the set of equations of motion for reduced Heisenberg operators for our problem reads (see Appendix \ref{A1}).
\begin{eqnarray}\label{ms}
\frac{d}{dt}\langle{\sigma_1^z}\rangle &=&
-\gamma_1(\langle\sigma_1^z\rangle+1)
-\gamma_{12}\langle\sigma_1^x\sigma_2^x\rangle
+\Lambda_{12}\langle\sigma_1^x\sigma_2^y\rangle 
-2G_0(t){\mathcal K}_0(t), 
\nonumber\\
\frac{d}{dt}\langle{\sigma_2^z}\rangle &=&
-\gamma_2(\langle\sigma_2^z\rangle+1)
-\gamma_{12}\langle\sigma_1^x\sigma_2^x\rangle
-\Lambda_{12}\langle\sigma_1^x\sigma_2^y\rangle  
-2G_1(t){\mathcal K}_1(t),
\nonumber\\
\frac{d}{dt}\langle{\sigma_1^x\sigma_2^x}\rangle &=&
-\frac{1}{2}(\gamma_1+\gamma_2)\langle\sigma_1^x\sigma_2^x\rangle 
+\frac{\gamma_{12}}{2}(\langle\sigma_1^z\rangle+\langle\sigma_2^z\rangle)
+\gamma_{12}\langle\sigma_1^z\sigma_2^z\rangle
+G_0(t){\mathcal Q}_0(t)+G_1(t){\mathcal Q}_1(t),
\nonumber\\ 
\frac{d}{dt}\langle{\sigma_1^x\sigma_2^y}\rangle &=&
-\frac{1}{2}(\gamma_1+\gamma_2)\langle\sigma_1^x\sigma_2^y\rangle
-\frac{\Lambda_{12}}{2}(\langle\sigma_1^z\rangle-\langle\sigma_2^z\rangle)
+iG_0(t)\mathcal{D}_0(t)-iG_1(t)\mathcal{D}_1(t),
\nonumber\\
\frac{d}{dt}\langle{\sigma_1^z\sigma_2^z}\rangle &=&
-(\gamma_1+\gamma_2)\langle\sigma_1^z\sigma_2^z\rangle
-\gamma_1\langle\sigma_2^z\rangle-\gamma_2\langle\sigma_1^z\rangle
+2\gamma_{12}\langle\sigma_1^x\sigma_2^x\rangle
-2[G_0(t){\mathcal Q}_1(t)+G_1(t){\mathcal Q}_0(t)],\nonumber\\
\end{eqnarray}
where  $\mathcal{K}_q(t)$,  $\mathcal{D}_q(t)$, and $\mathcal{Q}_q(t)$ are auxiliary functions defined in (\ref{au}) in the appendix \ref{A1}. A close look at (\ref{atoms2}) in the appendix \ref{A1} allows one to see that $\langle\sigma_1^x\sigma_2^x\rangle=\langle\sigma_1^y\sigma_2^y\rangle$ and $\langle\sigma_1^x\sigma_2^y\rangle=-\langle\sigma_1^y\sigma_2^x\rangle$. The constants appearing in (\ref{ms}) are given by
\begin{align}\label{gammai}
\gamma_{i}=2\pi\sum_\lambda\int d{\bf k} |g_{i}|^2\delta(\omega_k-\omega), 
\end{align}
\begin{align}\label{gamma12}
\gamma_{ij}=2\pi\sum_\lambda\int d{\bf k} g_ig^*_j\delta(\omega_k-\omega),  \ \ i\neq j,
\end{align}
\begin{align}\label{Lambda12}
\Lambda_{12}=2\sum_\lambda\int d{\bf k} \frac{g_1g^*_2}{\omega_k-\omega}.
\end{align}
The delta functions here express the fact that $g_i$ is essentially constant for
frequencies of interest which are centered around the atomic transition
frequency. The decay rates $\gamma_{i}$ are readily evaluated and the results are the well known free space atomic decay constants $\gamma_{i} = \tfrac{1}{3\pi} \left( \frac{\omega}{c} \right)^3 \frac{\mu_i^2}{\hbar \epsilon_0}$, where $\mu_i=|\mbox{\boldmath$\mu$}_{i}|$. The Lamb-shifts that would appear as a consequence of Cauchy principal values in $\gamma_{i}$ have been discarded since their contribution turned out to be insignificant compared to $\gamma_{i}$, $\gamma_{ij}$, and $\Lambda_{12}$. However, this is not the case for the cross terms, where the principal value gives rise to a  free space field-induced dipole-dipole coupling with magnitude $\Lambda_{12}$ (\ref{Lambda12}) which is comparable to $\gamma_{i}$ and $\gamma_{ij}$. This induced coupling has already been predicted for the spontaneous emission of originally non interacting atoms \cite{Natoms,Livro Ficek,Thiru}. To calculate integrals (\ref{gamma12}) and (\ref{Lambda12}), we set our coordinate system such that the direction of the $z$ axis coincides
with the direction the atomic separation vector $\mbox{\boldmath$r$}_{ij}=\mbox{\boldmath$r$}_{i}-\mbox{\boldmath$r$}_{j}$ \cite{Livro Ficek}, and we set the atomic dipoles parallel and polarized in the $xz$ plane as
$
\mbox{\boldmath$\mu_{i(j)}$}=\mu_{i(j)}[\sin\alpha,0,\cos\alpha],
$
where $\alpha$ is the angle between the dipole moments and $\mbox{\boldmath$r$}_{ij}$. The result is given by
\begin{eqnarray}\label{gam}
\gamma_{ij}&=&\frac{3}{2}\sqrt{\gamma_{1}\gamma_{2}}
\left\{
\sin^2\alpha\frac{\sin(\delta_{ij})}{\delta_{ij}}+(1-3\cos^2\alpha)\left[ \frac{\cos(\delta_{ij})}{(\delta_{ij})^2}-\frac{\sin(\delta_{ij})}{(\delta_{ij})^3}\right]
\right\},
\end{eqnarray}
and
\begin{eqnarray}\label{lam}
\Lambda_{12}&=&
\frac{3}{4}\sqrt{\gamma_{1}\gamma_{2}}
\sin^2\alpha\left\{
\frac{\cos(\delta_{ij})}{\delta_{ij}}
 +\frac{2}{\pi}\left[\frac{1}{\delta_{ij}^2}
-\frac{1}{\delta_{ij}}\mathcal{F}_1(\delta_{ij})\right]\right\} 
\nonumber\\ &&-\frac{3}{4}\sqrt{\gamma_{1}\gamma_{2}}(1-3\cos^2\alpha)
\left\{
\frac{\sin(\delta_{ij})}{\delta_{ij}^2}+
\frac{\cos(\delta_{ij})}{\delta_{ij}^3}-
\frac{2}{\pi}\left[\frac{1}{\delta_{ij}^2}\mathcal{F}_2(\delta_{ij})
+\frac{1}{\delta_{ij}^3}\mathcal{F}_1(\delta_{ij})\right]
\right\},
\end{eqnarray}
where $\delta_{ij}=2\pi r_{ij}/\lambda$, $\mathcal{F}_1(\theta)=\sin(\theta){\rm{Ci}}(\theta)-\cos(\theta){\rm{Si}}(\theta)$,
and $\mathcal{F}_2(\theta)=-\sin(\theta){\rm{Si}}(\theta)-\cos(\theta){\rm{Ci}}(\theta)$, with ${\rm{Si}}(\theta)$  [${\rm{Ci}}(\theta)$] sine [cosine] integral functions and $r_{ij}=|\mbox{\boldmath$r$}_{ij}|$. Here $\lambda$ is the wavelength associated with the atomic transition of angular frequency $\omega$. 

Finally, the temporal features of the pulse appear in (\ref{ms}) and (\ref{au}) through the functions $G_0(t)$ and $G_1(t)$ which have been defined as
\begin{align}
G_0(t)=\frac{\gamma_{10}}{\sqrt{\gamma_0}}\xi(t), \ G_1(t)=\frac{\gamma_{20}}{\sqrt{\gamma_0}}\xi(t),
\end{align}
where $\xi(t)$ is related to the spectral function $f(\omega_k)$ in the initial field state (\ref{1p}) through the transform (\ref{t}). Now we have all the tools necessary to analyze the physical properties of the atomic system under single photon excitation in free space, including the single-photon induced build up of quantum correlations between the atoms in free space.

\section{Quantum Correlations} 
Quantum mechanics allows the existence of correlations providing quantum information resources not present in classical systems. One example is  quantum entanglement which is defined as the type of correlation that can not be created using local operations and classical communications. 
We aim now to study the arising of this kind of quantum correlation between the atoms due to their interaction with a single photon in free space.
Since we are dealing with an open system, and the atomic state is certainly mixed, we will consider here the entanglement of formation which is a suitable entanglement measure for such cases. For two two-level systems, this entanglement measure is given by \cite{Woo}
\begin{eqnarray}\label{eof}
EoF=h[(1+\sqrt{1-C^2})/2],
\end{eqnarray}
where $h(x)=-x\log_2x-(1-x)\log_2(1-x)$ and $C$ is the concurrence. The solution of (\ref{ms}) led to an atomic density operator which falls into the class of X states \cite{X} for which concurrence and then entanglement of formation can be readily obtained. From the density operator matrix elements in computational basis, concurrence of an X state can be calculated as 
\begin{eqnarray}
C=2\max\{0,|\rho_{23}(t)|-\sqrt{\rho_{11}(t)\rho_{44}(t)},|\rho_{14}(t)|-\sqrt{\rho_{22}(t)\rho_{33}(t)}\}.
\end{eqnarray}

\begin{figure}[ht]
 \centering\includegraphics[width=0.45\columnwidth]{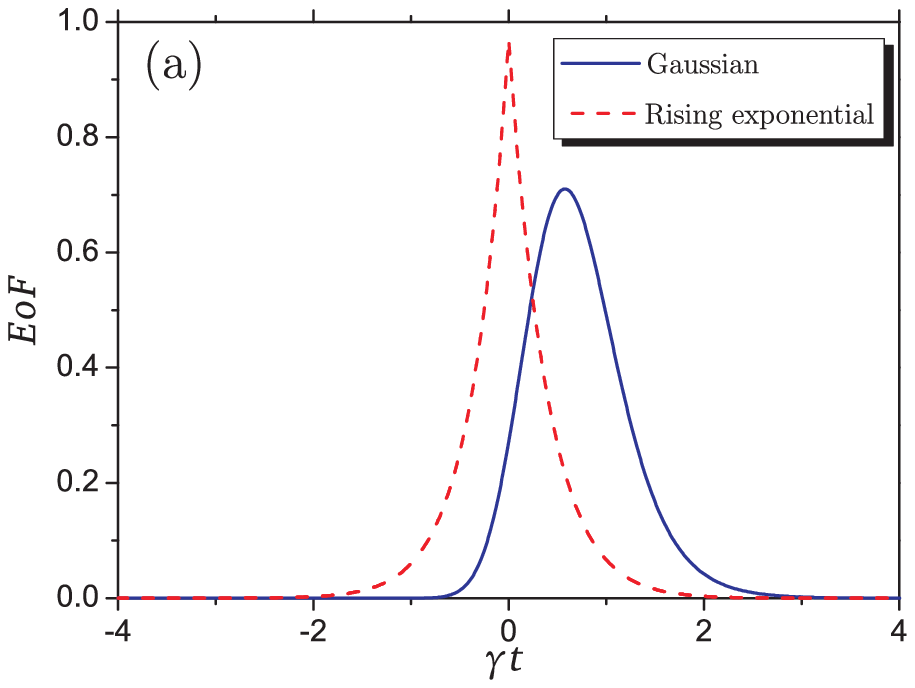} \centering\includegraphics[width=0.45\columnwidth]{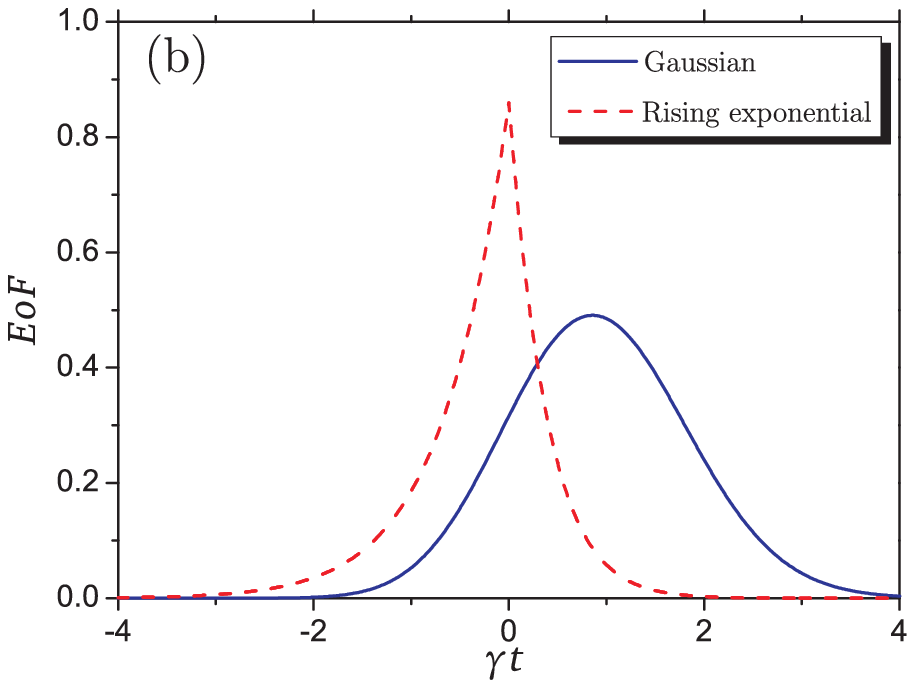}
 \caption{Entanglement dynamics for initially unexcited atoms in interaction with a single photon pulse for different atomic separations and pulse shapes. In (a) the parameters used for the Gaussian pulse were $(r_{12}/\lambda=0.2,\Omega/\gamma=2.55)$ and for the rising exponential pulse $(r_{12}/\lambda=0.2, \Omega/\gamma=1.77)$. In (b) the parameters used for the Gaussian and rising pulse were $(r_{12}=0.2, \Omega/\gamma=1.0)$.}
 \label{fig1}
\end{figure}

\begin{figure}[ht]
 \centering\includegraphics[width=0.45\columnwidth]{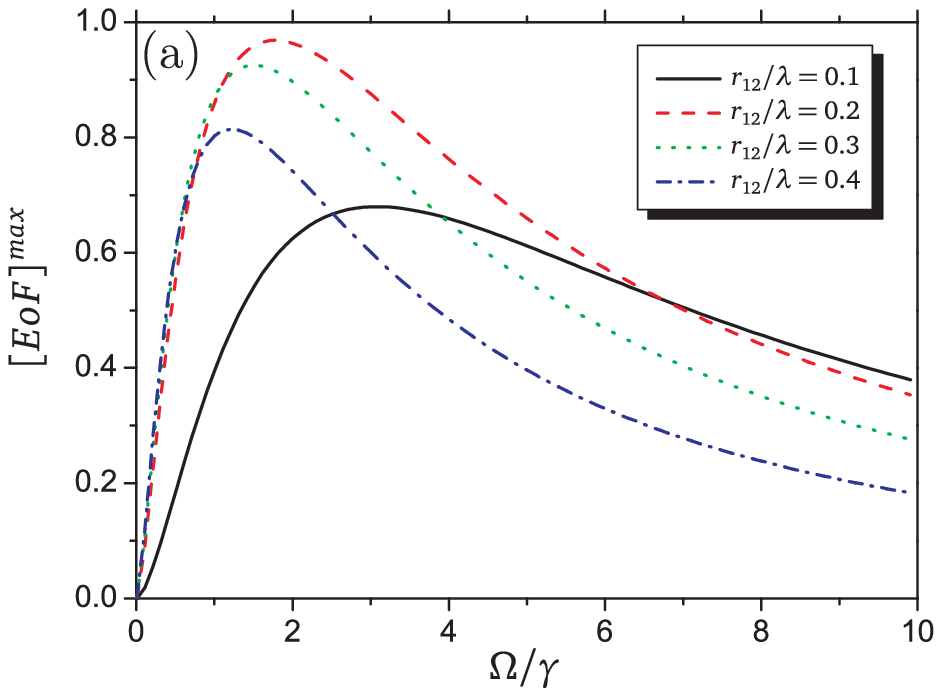} \centering\includegraphics[width=0.45\columnwidth]{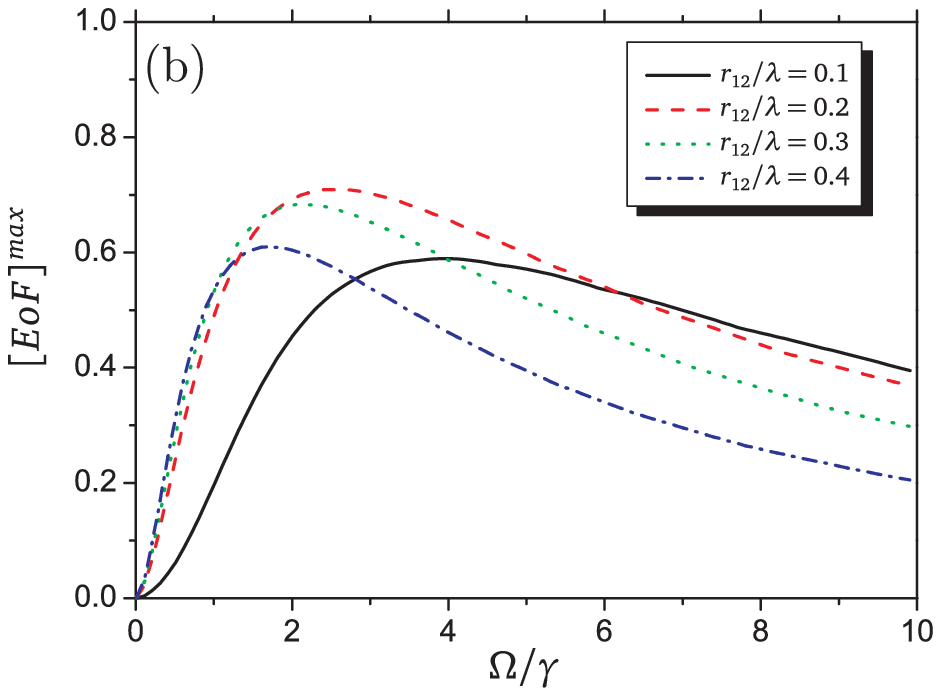}
 \caption{Maximal entanglement for several atomic distances as a function of the normalized pulse bandwidth $\Omega/\gamma$. (a) Rising exponential pulse (b) Gaussian pulse.}
 \label{fig2}
\end{figure}

In the following plots, we will keep $\alpha=\pi/2$, $\gamma_1=\gamma_2=\gamma$ and $\mbox{\boldmath$r$}_0$ at the midpoint between $\mbox{\boldmath$r$}_1$ and $\mbox{\boldmath$r$}_2$ in order not to favor any of the two atoms. The initial system state in $t\rightarrow -\infty$ was $|\Psi\rangle=|g_1,g_2,1_p\rangle$, as explained before. Figure \ref{fig1} shows a typical dynamics for the entanglement between the atoms in interaction with a single photon in a rising exponential pulse or a Gaussian pulse. Each choice of $\Omega/\gamma$ corresponds to an optimum value regarding the peak of the entanglement curve for a fixed atomic separation. We can see that the rising exponential pulse is able to create more entanglement between the atoms than the Gaussian pulse. This might be related to the fact that rising exponentials correspond to the reversed dipole waves which perfectly excites a single atom, but the analogy is not complete here due to the presence of a second atom which couples to the first through the quantized electromagnetic field. Also, by comparing (a) and (b) in figure \ref{fig1}, one can notice the great sensitivity to variation of pulse  bandwidths $\Omega$ in (\ref{r}) and (\ref{ga}). The whole atomic response to the single photon pulse must also change strongly depending on the atomic relative separation $r_{12}$ through the collective damping rate (\ref{gam}) and  free space field-induced coupling constant (\ref{lam}). Figure \ref{fig2} illustrates this point by showing the maximally achieved entanglement (peak value) for several atomic separations $r_{12}/\lambda$ as a function of the pulse bandwidth $\Omega/\gamma$.  One can see from figure \ref{fig2} that it is indeed possible to optimize $\Omega$ in order to maximize the peak value of entanglement for a fixed relative distance $r_{12}$ between the atoms, just like done in Figure \ref{fig1}. 


Our calculation provides a quantitative and complete quantum electrodynamic description of the interaction between two two-level atoms and one single photon in free space. From this model it is possible to understand that the building up of entanglement between the atoms can not be solely explained using the qualitative picture which asserts that the atoms get entangled because they ``see'' the same photon in the \textit{absorption process}. The situation is quite richer than that as exemplified in figure \ref{fig2}. Had the building up of entanglement been only dependent on this fact, the closer the atoms the bigger the entanglement would have become. However, plots in figure \ref{fig2} clearly reveal that there is an optimum atomic separation for each pulse shape. For example, by fixing $\Omega/\gamma\approx 4.0$, one can see that $r_{12}/\lambda=0.2$ is more favorable to entanglement generation (maximum entanglement) than $r_{12}/\lambda=0.1$. This is a consequence of the fact that the atomic coupling (\ref{lam}) and the collective spontaneous emission (\ref{gam}) depend on the atomic distance in an oscillatory manner, and the equations of movement depend on the pulse shape. Consequently there is a competition among several factors which determines the atomic response. All these factors can be completely understood only within a full quantum electrodynamic calculation such as the one presented here.

It is also important to contrast our findings and previous work on the generation of atomic entanglement by detection of indistinguishable single-photon \textit{emissions} \cite{int}, especially the work by Cabrillo \textit{et al} \cite{cab}. In this case, entanglement between the atoms is generated by interference effects and
\textit{state projection} accompanying a measurement. State projection is an essential ingredient in the scheme. 
In our study, there is no measurement whatsoever and our main goal is to show an alternative way to create entanglement among atoms which is due to the electromagnetic pulse and vacuum induced couplings. In our scheme the atoms are originally in the ground state while in the approach showed in \cite{cab} the atoms are initially prepared in the excited state by \textit{two} classical laser fields. Consequently, there is a chance of measuring both atoms in excited state and this clearly distinguishing our proposal from theirs since here this joint probability is always zero. This follows from the fact that there is just one single photon in the driving quantum pulse. It may be argued that our proposal is not quite robust since spontaneous emission eventually destroys all the entanglement. However, it is precisely the emission of a photon by the entangled pair that can be used to propagate information between quantum nodes. The idea is to control the pulses optimally in the time window where correlations can be transferred from distant nodes by photons. This idea o quantum memory in free space is under current investigation \cite{Scarani2}.

It is worthwhile to compare the methods normally used in CQED to the ones available in these free space quantum optical setups. In CQED, atomic states are usually engineered by suitably choosing the initial state and the time of interaction with monochromatic fields. Here, the idea is to control the spectral properties of the pulse in interaction with atoms initially prepared in some fiducial initial state. We think that the results presented here may serve as a benchmark for new investigations about quantum state engineer and quantum communication in free space by using single photons as a tool. For example, it would be interesting to see whether or not a general state $|\phi\rangle=\alpha|e_1,g_2\rangle-\beta|g_1,e_2\rangle$ may be generated from the single photon pulse of equation (\ref{1p}) by suitably choosing $f(\omega_k)$. We believe that a generalization of the time reversal argument used in \cite{perf_exc} to find the pulse state leading to a perfect $\pi-$pulse would allow one to solve this problem, but we have not tested this conjecture.

\section{Conclusions} 
We presented a fully quantized model for describing the interaction of single photons with atoms in free space. We showed that the free space correlated decay and dipole-dipole coupling between the atoms in conjunction with the single photon pulse can induce atomic quantum correlations.
All features displayed by the atomic entanglement are strongly dependent on the atomic relative positions and the pulse profile, so that it is possible to achieve high control of the entanglement dynamics by carefully setting the system and pulse parameters.  We believe our results can be important for quantum communication where the build up and distribution of atomic correlations by photon pulses is an essential ingredient. One possible future route is the use of a second pulse to redistribute the entanglement generated by the first pulse as predicted here. This might follow from results started in \cite{Scarani2}, where the idea of quantum memory with a single atom in a half cavity is proposed. 

Our proposal has been made in the context of photonic qubits in free space but it can also be relevant to circuit quantum electrodynamics where strong extinction of the transmitted power in a 1D transmission line coupled to an artificial atom has recently been achieved \cite{sci}. In fact, cooperative effects, just like the ones presented here in the context of atomic physics, are under theoretical and experimental investigation in circuit QED systems \cite{cqed}, and our results may find applications in these systems as well. 
\begin{acknowledgments}
J.P.S acknowledges FAPESP (Fundação de Amparo \`a Pesquisa do Estado de São Paulo) Grant No. 2011/09258-5, Brazil. F.L.S. acknowledges partial support from CNPq (Grant No. 308948/2011-4) and the Brazilian National Institute of Science and Technology of Quantum Information (INCT-IQ).
\end{acknowledgments}
\appendix

\section{Equations of Motion}\label{A1}

In this appendix we present details about the derivation of the system of equations (\ref{ms}). From formal integration of the Heisenberg equation for $a^{}_{{\bf k},\lambda}(t)$ obtained with Hamiltonian (\ref{H}) one obtains
\begin{align}
a^{}_{{\bf k},\lambda}(t)&=a^{}_{{\bf k},\lambda}(t_0)+
\int_{t_0}^{t}dt'g^*_{1}\sigma_1^{-}(t')e^{i(\omega_k-\omega)t'}
+\int_{t_0}^{t}dt'g^*_{2}\sigma_2^{-}(t')e^{i(\omega_k-\omega)t'}. \label{bosonic1}
\end{align}
On the other hand, a general atomic operator ${\mathcal O}$ evolving according to Hamiltonian (\ref{H})
will obey
\begin{align}\label{heisembergO}
\frac{d}{dt}{\mathcal O(t)}=
\sum_\lambda\int d{\bf k}
[(
g_{1}[\sigma_1^+(t),{\mathcal O(t)}]+g_{2}[\sigma_2^+(t),{\mathcal O(t)}])a^{}_{{\bf k},\lambda}(t) e^{-i(\omega_{k}-\omega)t}
\nonumber\\-
a^{\dag}_{{\bf k},\lambda}(t)(
g^*_{1}[\sigma_1^{-}(t),{\mathcal O}(t)]+g^*_{2}[\sigma_2^{-}(t),{\mathcal O}(t)]) e^{i(\omega_{k}-\omega)t}].
\end{align}
Even though atomic and field operators commute when taken at the same time, the choice of order here may simplify terms in further developments \cite{cohen}. We found it convenient to write the operator $a_{{\bf k},\lambda}(t)$ on the right and the operator $a^{\dag}_{{\bf k},\lambda}(t)$ on the left just like used in another context in \cite{cohen} on page 394. By substituting (\ref{bosonic1}) in (\ref{heisembergO}) and changing the integration variable to $\tau=t-t'$, one gets
\begin{align}\label{tau}
\frac{d}{dt}{\mathcal O(t)}&=
\sum_{i=1}^2\sum_\lambda\int d{\bf k}
\left[
g_{i}[\sigma_i^+(t),{\mathcal O(t)}]a^{}_{{\bf k},\lambda}(t_0) e^{-i(\omega_{k}-\omega)t}
-
a^{\dag}_{{\bf k},\lambda}(t_0)
g^*_{i}[\sigma_i^{-}(t),{\mathcal O}(t)] e^{+i(\omega_{k}-\omega)t}\right]
\nonumber\\
&+\sum_{i,j=1}^2\sum_\lambda\int d{\bf k}
\left[
g_{i}^*g_{j}[\sigma_j^+(t),{\mathcal O(t)}]\int_{0}^{t-t_0}d\tau \sigma_i^{-}(t-\tau)e^{-i(\omega_k-\omega)\tau}-
\int_{0}^{t-t_0}d\tau \sigma_i^{+}(t-\tau)e^{i(\omega_k-\omega)\tau}
g_{i}g_{j}^*[\sigma_j^{-}(t),{\mathcal O}(t)]\right].
\end{align}
From now on, one must take care in order to not change the order in which atomic and field operators appear since now they appear evaluated at different times and then they do not commute. The last two terms in (\ref{tau}) appear in the form of a integral over $\tau$ of $\sigma_{i}^{+(-)}(t-\tau)$ multiplied by the function
\begin{align}\label{integral}
f(t)=\sum_\lambda\int d{\bf k}
g^*_{i}g_{j}e^{-i(\omega_k-\omega)\tau}
\end{align}
or its conjugate. The frequencies $\omega_k$ of the field modes cover a very wide range and $g_{i}g^*_{j}$ varies slowly with respect to $\omega_k$. It follows that the set of oscillating exponentials of $f(t)$ interfere destructively when $\tau$ increases starting from $0$, so that $f(t)$ becomes negligible as soon as $\tau\gg\tau_c$, where $\tau_c$ is the correlation time of the vacuum fluctuations.
In addition, $\sigma_{i}^{+(-)}(t-\tau)$ varies much more slowly with $\tau$, over a time scale $T_R\gg\tau_c$, where $T_R$ is the damping time of the atoms. We can then neglect the variation with $\tau$ of $\sigma_{i}^{+(-)}(t-\tau)$ as compared that of $f(t)$ in the integral (\ref{tau}), and replace $\sigma_{i}^{+(-)}(t-\tau)$ by $\sigma_{i}^{+(-)}(t)$ which may then be removed from the integral. Assuming $t-t_0\gg\tau_c$, we get in this way
\begin{align}\label{atoms}
\frac{d}{dt}{\mathcal O}=
\sum_{i=1}^2\sum_\lambda\int d{\bf k}
\Bigl[
g_{i}[\sigma_i^+,{\mathcal O}]
a^{}_{{\bf k},\lambda}(t_0)e^{-i(\omega_{k}-\omega)t}
+
g^*_{i}a^{\dag}_{{\bf k},\lambda}(t_0)
[{\mathcal O},\sigma_i^{-}]
e^{+i(\omega_{k}-\omega)t}\Bigl]
\nonumber\\ 
+\sum_{i,j=1}^2\sum_\lambda\int d{\bf k}\Bigl[
g_{i}g^*_{j}
\chi^{(-)}[\sigma_i^+,{\mathcal O}]\sigma_j^{-}(t)
+g^*_{i}g_{j}
\chi^{(+)}\sigma^+_j(t)[{\mathcal O},\sigma_i^{-}]
\Bigl]
\end{align}
with
\begin{align}
\chi^{(\pm)}\equiv\pi\delta(\omega_k-\omega)\pm i\frac{{\mathcal P}}{\omega_k-\omega},
\end{align}
where ${\mathcal P}$ indicate the Cauchy Principal Value. In order to obtain (\ref{ms}), one needs to get (\ref{atoms}) projected onto the initial state
$|\psi_1\rangle=|g_1,g_2,1_p\rangle$, where
\begin{align}\label{initial}
|1_p\rangle=\sum_\lambda\int d{\bf k} g^*_{0;{{\bf k},\lambda}}f(\omega_k)a_{{\bf k},\lambda}^\dag(t_0)|0\rangle.
\end{align}
All calculations are then performed in the Weisskopf-Wigner approximation
where it is assumed that the coupling constant $g_i$ is constant for frequencies of interest centered around the atomic transition
frequency $\omega$ such that
\begin{align}\label{G}
\left[\sum_\lambda\int d{\bf k} g_{i}a_{{\bf k},\lambda}(t_0)e^{-i(\omega_k-\omega)t}\right]
|1_p\rangle=G_{i-1}(t)|0\rangle,
\end{align}
where 
\begin{align}
G_{i-1}(t)=\frac{\gamma_{i0}}{\sqrt{\gamma_0}}\xi(t),
\end{align}
and $\gamma_{i0}$, $\gamma_0$ and $\xi(t)$ were defined in (\ref{gamma12}), (\ref{gammai}) and (\ref{t}). If ${\mathcal O}$ is Hermitian, we have from (\ref{atoms}) that
\begin{align}\label{atoms2}
\frac{d}{dt}\langle\psi_1|{\mathcal O}|\psi_1\rangle=\sum_{i=1}^2
2{\rm Re}\Bigl[G_{i-1}\langle \psi_1|[\sigma_i^+,{\mathcal O}]|\psi_0\rangle\Bigr]
+\sum_{i,j=1}^2\sum_\lambda\int d{\bf k} \ 
2{\rm Re}\Bigl[g_{i}g^*_{j}\chi^{(-)}\langle\psi_1|[\sigma_i^+,{\mathcal O}]\sigma_j^{-}(t)|\psi_1\rangle\Bigr],
\end{align}
from which (\ref{ms}) can be readily obtained. The auxiliary functions $\mathcal{K}_q(t)$, $\mathcal{D}_q(t)$, and $\mathcal{Q}_q(t)$ present in (\ref{ms}) have been defined as
\begin{align}\label{au}
\mathcal{K}_q(t)&=2\,{\rm{Re}}\langle\psi_0|\sigma_{q\oplus 1}^-|\psi_1\rangle\,\,\,(q=0,1),\nonumber\\
\mathcal{D}_0(t)&=2i\,{\rm{Im}}\langle\psi_0|\sigma_1^z\sigma_2^-|\psi_1\rangle, \nonumber\\
\mathcal{D}_1(t)&=2i\,{\rm{Im}}\langle\psi_0|\sigma_1^-\sigma_2^z|\psi_1\rangle, \nonumber\\ \mathcal{Q}_0(t)&=2\,{\rm{Re}}\langle\psi_0|\sigma_1^z\sigma_2^-|\psi_1\rangle, \nonumber\\
\mathcal{Q}_1(t)&=2\,{\rm{Re}}\langle\psi_0|\sigma_1^-\sigma_2^z|\psi_1\rangle,
\end{align}
with $|\psi_1\rangle=|g_1,g_2,1_p\rangle$ and $|\psi_0\rangle=|g_1,g_2,0_p\rangle$. The equations of motion for the auxiliary system can be obtained from
\begin{align}
\frac{d}{dt}{\langle\psi_0|\mathcal O|\psi_1\rangle}=
\sum_{i=1}^2
\Bigl[
G_{i-1}\langle\psi_0|[\sigma_i^+,{\mathcal O}]|\psi_0\rangle
+\sum_{i,j=1}^2\sum_\lambda\int d{\bf k}
g_{i}g^*_{j}
\chi^{(-)}\langle\psi_0|[\sigma_i^+,{\mathcal O}]\sigma_j^{-}(t)|\psi_0\rangle 
+g^*_{i}g_{j}
\chi^{(+)}\langle\psi_0|\sigma^+_j(t)[{\mathcal O},\sigma_i^{-}]|\psi_0\rangle
\Bigl],
\end{align}
which follows from (\ref{atoms}) and (\ref{G}). It results in the system
\begin{eqnarray}\label{au}
\dot{\mathcal{J}_q}(t)&=&-\frac{1}{2}\gamma_q\mathcal{J}_q(t)+\frac{1}{2}\gamma_{12}\mathcal{D}_q(t)-i\frac{\Lambda_{12}}{2}\mathcal{Q}_q \nonumber\\
\dot{\mathcal{K}_q}(t)&=&-\frac{1}{2}\gamma_q\mathcal{K}_q(t)+\frac{1}{2}\gamma_{12}\mathcal{Q}_q(t)-i\frac{\Lambda_{12}}{2}\mathcal{D}_q-2G_q \nonumber\\
\dot{\mathcal{D}_q}(t)&=&-(\frac{1}{2}\gamma_{q\oplus 1}+\gamma_q)\mathcal{D}_q(t)
-\gamma_{12}\mathcal{D}_{q\oplus 1}(t) -\frac{1}{2}\gamma_{12}\mathcal{J}_q(t)-\gamma_q\mathcal{J}_{q\oplus 1}(t)-\frac{i\Lambda_{12}}{2}\mathcal{K}_q \nonumber\\
\dot{\mathcal{Q}_q}(t)&=&-(\frac{1}{2}\gamma_{q\oplus 1}+\gamma_q)\mathcal{Q}_q(t)
-\gamma_{12}\mathcal{Q}_{q\oplus 1}(t)-\frac{1}{2}\gamma_{12}\mathcal{K}_q(t)-\gamma_q\mathcal{K}_{q\oplus 1}(t)-\frac{i\Lambda_{12}}{2}\mathcal{J}_q+2G_{q\oplus 1},
\end{eqnarray} 
with $\oplus$ corresponding to addition modulo 2 ($q$ assumes values in $\{0,1\}$) and $\mathcal{J}_q(t)=2i\,{\rm{Im}}\langle\psi_0|\sigma_{q}^-|\psi_1\rangle$. 
%

\end{document}